\documentstyle[preprint,aps,eqsecnum,epsfig]{revtex}

\newcommand{\beq}{\begin{equation}}
\newcommand{\eeq}{\end{equation}}
\newcommand{\beqs}{\begin{eqnarray}}
\newcommand{\eeqs}{\end{eqnarray}}

\def\mathcal{\cal}
\def\mathrm{\rm}
\def\textbf{\bf}
\def\textit{\it}
\def\lesssim{\ \rlap{\raise 3pt \hbox{$<$}}{\lower 3pt \hbox{$\sim$}}\ }
\def\gtrsim{\ \rlap{\raise 3pt \hbox{$>$}}{\lower 3pt \hbox{$\sim$}}\ }
\def\vev#1{\langle #1 \rangle}

\def\Re{{\cal R}e}
\def\Im{{\cal I}m}

\def\ifmath#1{\relax\ifmmode #1\else $#1$\fi}
\def\ifmath{}

\begin{document}
\draft

{\tighten
\preprint{\vbox{\hbox{WIS-00/46/JAN-DPP}
                \hbox{hep-ph/0001037}}}
\title{~\\~\\The $K_L\to\pi^0\nu\bar\nu$ Decay in Models
of Extended Scalar Sector}

\author{Gilad Perez}
\address{ \vbox{\vskip 0.truecm}
  Department of Particle Physics,
  Weizmann Institute of Science, Rehovot 76100, Israel }
\maketitle

\begin{abstract}
We calculate new contributions to the
$K_L\to\pi^0\nu\bar\nu$ decay
in models where neutrino Majorana masses require
an extension of the scalar sector.
First, we study a model where the neutrino mass is
induced by the vacuum expectation value of an $SU(2)$-triplet
scalar. Second, we study the Zee model where the Majorana
mass comes from one loop diagrams involving a singly charged,
$SU(2)$-singlet scalar. In both models, the Yukawa couplings
that involve the new scalar and the neutrinos could be
of order one. We find, however, that the contributions
to the $K_L\to\pi^0\nu\bar\nu$ decay mediated
by the new scalars depend on the neutrino
masses rather than the Yukawa couplings and are, therefore,
negligibly small.
\end{abstract}
} %end tighten

\newpage

\section{Introduction}
Within the standard model (SM) the
$K_{L}\rightarrow \pi ^{0}\nu \bar{\nu }$ decay is known to be
a CP violating (CPV) process to a very good approximation
\cite{LIT,BI}, and subject to a clean theoretical interpretation
\cite{Bur}. In the SM, CP conserving (CPC) contributions
are chirally suppressed and smaller by many orders of magnitude than
the CPV ones \cite{BI}. In a previous work \cite{GP}, we calculated the CPC
contributions that arise when the SM Lagrangian is extended to include
neutrino mass terms. These contributions are known to be chirally
enhanced, but we found that they are suppressed
by a factor of order $m_\nu^2/m_W^2$ which makes them negligibly small.
In this work we study models where, in addition to neutrino mass terms,
there are new light scalars. The question that we ask is whether
the fact that such scalars can have Yukawa couplings of order one
to neutrinos allows for a situation where the new contributions
are chirally enhanced while avoiding the $m_\nu^2/m_W^2$
suppression factor.

Specifically, we consider the following two models:

1. Neutrino masses are related to the vacuum expectation value of
an $SU(2)$-triplet scalar. We present the model and calculate the
new contributions to the $K_L\to\pi^0\nu\bar\nu$ decay in section II.

2. Neutrino masses are related to loop diagrams involving a singly charged
$SU(2)$-singlet and an additional $SU(2)$-doublet scalar.
We present the model and calculate the
new contributions to the $K_L\to\pi^0\nu\bar\nu$ decay in section III.

A summary of our conclusions is given in section IV.

\section{An $SU(2)$ Triplet Scalar}

\subsection{The Model}

We consider the SM Lagrangian with the addition of an SU(2)
triplet Higgs field ($\Delta$).
The most general scalar potential is given by \cite{Small}:
\begin{eqnarray}
V(H,\Delta) &=& \frac{\lambda_H}{2}({H}_i^\dagger H^i)^2
+\frac{\lambda_{\Delta}}{2}({\Delta}_t^\dagger \Delta^t)^2
+\lambda H_i^\dagger H^i \Delta_t^\dagger \Delta^t
+\widetilde{\lambda}
{H^i}^\dagger \sigma_{ij}^q  H^j
{\Delta^r}^\dagger t_{rs}^q \Delta^s
\nonumber \\
&+& m(\bar{\Delta}^0 H^0 H^0+\sqrt{2}\Delta^- H^+ H^0
+\Delta^{--} H^+ H^+)+
\mu_H^2 {H}_i^\dagger H^i
+\mu_\Delta^2 {\Delta}_t^\dagger {\Delta}^t+h.c
\, ,
\label{scalpot}
\end{eqnarray}
where $\Delta^t = [\Delta^{++},\Delta^+,\Delta^0]$,
$H^i = [H^+,H^0]$ and
$\sigma,t$ are are SU(2) generators in the \linebreak
$J=1/2,\ 1$ representations, respectively.
The $m$-term in (\ref{scalpot}) breaks the global lepton number symmetry
($L$) so that the phenomenologically unacceptable \cite{GGN,EPJ}
Majoron \cite{Gel} is avoided.
For simplicity, we assume that the couplings in (\ref{scalpot})
conserve CP. As concerns the VEVs,
$\vev{H^0}\equiv \frac{v_H}{\sqrt{2}}$ and
$\vev{\Delta^0}\equiv \frac{v_\Delta}{\sqrt{2}}$,
we assume for simplicity that they
are both real and we take into account the constraints from the
$\rho$ parameter \cite{Gel,Pec}:
\begin{equation}
\frac{v_\Delta}{v_H}\lesssim 10^{-2}
\label{small}
\, .
\end{equation}

In order to calculate the contribution to the decay, the
scalar mass eigenstates and mixing angles should be
identified. With the assumption that the
scalar potential $V(H,\Delta)$ (\ref{scalpot}) is real, the
imaginary and the real parts of the neutral scalars remain
unmixed.
For the CP even fields, $\sqrt{2}\Re(H^0)$ and
$\sqrt{2}\Re(\Delta^0)$, we have the
following mass matrix \cite{Small}~(we neglect terms that are
higher order in $v_\Delta/v_H$ and assume that the $\lambda$'s
are positive and of order one):
\begin{equation}
M_{\cal R}^2 =
v_H^2 \pmatrix{\lambda_H &
\frac{\sqrt{2}}{v_H}(\hat\lambda \frac{v_\Delta}{\sqrt{2}} +m) \cr
\frac{\sqrt{2}}{v_H}(\hat\lambda \frac{v_\Delta}{\sqrt{2}} +m) &
-\frac{m}{\sqrt{2}v_\Delta}}
 \, , \label{MrH0Del0}
\end{equation}
where $\hat\lambda \equiv \lambda+\widetilde{\lambda}$.
The eigenvectors of $M^2_{\cal R}$
are:
\begin{eqnarray}
&{\xi}_1& = \Re(H^0)\cos\gamma+\Re(\Delta^0)\sin\gamma
\nonumber \\
&{\xi}_2& = -\Re(H^0)\sin\gamma+\Re(\Delta^0)\cos\gamma
\, , \label{xis}
\end{eqnarray}
where
\begin{eqnarray}
\tan \gamma &=&
-\frac{v_H}{2v_\Delta}\left(\hat\lambda
+\frac{\sqrt{2}m}{v_\Delta}\right)^{-1}
\left[\lambda_H + \frac{m}{\sqrt{2}v_\Delta}
-\sqrt{\left(\lambda_H+\frac{m}{\sqrt{2}v_\Delta}\right)^2
+4\frac{v_\Delta^2}{v_H^2}
\left(\hat\lambda+\frac{\sqrt{2}m}{v_\Delta}\right)^2}\right]
\nonumber \\
&\approx& 2\frac{v_\Delta}{v_H}
\left(\frac{\hat\lambda+\frac{\sqrt{2}m}{v_\Delta}}
{2\lambda_H+\frac{\sqrt{2}m}{v_\Delta}}\right)\ll 1
 \, . \label{tang}
\end{eqnarray}
In deriving the inequality in eq.~(\ref{tang}) and below
we assume that there are no fine-tuned cancellations among
the independent parameters of the scalar sector.
The eigenvectors correspond to the following mass
eigenvalues:
\begin{eqnarray}
m^2_{\xi_{1,2}} &=&
\frac{v_H^2}{2} \left[\lambda_H -\frac{m}{\sqrt{2}v_\Delta}
\pm \sqrt{\left(\lambda_H+\frac{m}{\sqrt{2}v_\Delta}\right)^2
+4\frac{v_\Delta^2}{v_H^2}
\left(\hat\lambda+\frac{\sqrt{2}m}{v_\Delta}\right)^2} \right]
\nonumber \\
&\approx& v_H^2
\left[ \lambda_H+\frac{v_\Delta^2}{v_H^2}
\frac{\left(\hat\lambda+\frac{\sqrt{2}m}{v_\Delta}\right)^2}
{\lambda_H+\frac{m}{\sqrt{2}v_\Delta}}
,-\frac{m}{\sqrt{2}v_\Delta}+O\left(m\frac{v_\Delta^2}{v_H^2}\right)
\right]
 \, . \label{mrHL}
\end{eqnarray}
For the CP odd fields, $\sqrt{2}\Im(H^0)$ and
$\sqrt{2}\Im(\Delta^0)$,
we have the following mass matrix:
\begin{equation}
M_{\cal I}^2 =
\sqrt{2}m \pmatrix{-2v_\Delta &-v_H \cr
-v_H & -\frac{v_H^2}{2v_\Delta}}
 \, . \label{MiH0Del0}
\end{equation}
The eigenvectors of $M^2_{\cal I}$
are:
\begin{eqnarray}
G^0&=& \Im(H^0)\cos\eta+\Im(\Delta^0)\sin\eta
\nonumber \\
J^0&=& -\Im(H^0)\sin\eta+\Im(\Delta^0)\cos\eta
\, , \label{A_I}
\end{eqnarray}
where
\begin{equation}
|\tan \eta| = \left|\frac{2v_\Delta}{v_H}\right|\ll 1
 \, . \label{tanet}
\end{equation}
We learn that mixing in the CP-odd sector is very small and we
neglect it from here on.
The eigenvectors correspond to the following mass
eigenvalues \cite{Small}:
\begin{equation}
m_{G^0,J^0}^2 =
\left[0,-\frac{mv_H^2}{\sqrt{2}v_\Delta}\right]
 \, , \label{mim}
\end{equation}
where the massless component $G^0 \sim \sqrt{2}\Im(H^0)$
corresponds to the unphysical Goldstone boson which is eaten
by the $Z$ boson, while $J^0 \sim \sqrt{2}\Im(\Delta^0)$
corresponds to the would-be Majoron. As expected its squared
mass is proportional to the explicit
lepton number violating parameter $m$.

For the singly charged scalars, $H^\pm$ and $\Delta^\pm$,
we have the following mass matrix:
\begin{equation}
M_{C}^2 =
\frac{v_H}{\sqrt{2}}
(\widetilde{\lambda}\frac{v_\Delta}{\sqrt{2}}+m)
\pmatrix{-2\frac{v_\Delta}{v_H} & \sqrt{2} \cr
\sqrt{2} &
-\frac{v_H}{v_\Delta}}
 \, . \label{MchargeDelH}
\end{equation}
The eigenvectors of $M_{C}^2$ are:
\begin{eqnarray}
&G^\pm& = H^\pm \cos\eta' +\Delta^\pm \sin\eta'
\nonumber \\
&\xi^\pm& = -H^\pm \sin\eta'+\Delta^\pm \cos\eta'
\, , \label{G_xi_pm}
\end{eqnarray}
where
\begin{equation}
|\tan \eta'| = \left|\frac{\sqrt{2}v_\Delta}{v_H}\right|
\ll 1
 \, . \label{tanet'}
\end{equation}
We learn that mixing in the charged sector is very small and we
neglect it from here on.
The eigenvectors correspond to the following mass
eigenvalues:
\begin{equation}
m_{G^\pm,\xi^\pm}^2 =
\left[0,-\frac{v_H^2}{\sqrt{2}}
\left(\frac{\widetilde{\lambda}}{\sqrt{2}}
+\frac{m}{v_\Delta}\right)\right]
 \, , \label{mcharge}
\end{equation}
where the massless component $G^\pm \sim H^\pm$
corresponds to the unphysical charged Goldstone boson which
is eaten by the $W^\pm$ bosons, while $ \xi^\pm\sim
\Delta^\pm$ corresponds to a new physical charged scalar.

Neutrino Majorana masses are
induced via the following interaction terms:
\begin{equation}
{\mathcal{L}}_{\nu \nu }=f_{mn}(L_{Li}^{m})_{\alpha }
\tau_{\alpha \beta }^{2}\sigma_{ir}^{2}\sigma
_{rj}^{t}(L_{Lj}^{n})_{\beta }\Delta ^t
\label{tripmass}
\, ,
\end{equation}
where $\tau$ is an SU(2) generator in the
$J=1/2$ spinor representation.
For simplicity, we take $f_{mn}=f_m\delta_{mn}$ with
$f_m$ real. In this way, we avoid unnecessary complications
related to flavor mixing and CP violation in the neutrino
sector (see discussion in ref.~\cite{GP}).
The interaction term (\ref{tripmass}) induces neutrino masses:
\begin{equation}
\label{tripmassterm}
(m^{Maj}_\nu)_{i}= f_{i}{\langle\Delta^0\rangle}
\end{equation}
with $i=1,2,3 \,$.
In addition, it generates new contributions to the
$K_{L}\rightarrow \pi ^{0}\nu \bar{\nu}$ decay.

\subsection{The $K_L\to\pi^0\nu\bar\nu$ Decay Rate}

\begin{figure}[hbct]
\begin{center}
\mbox{\epsfig{figure=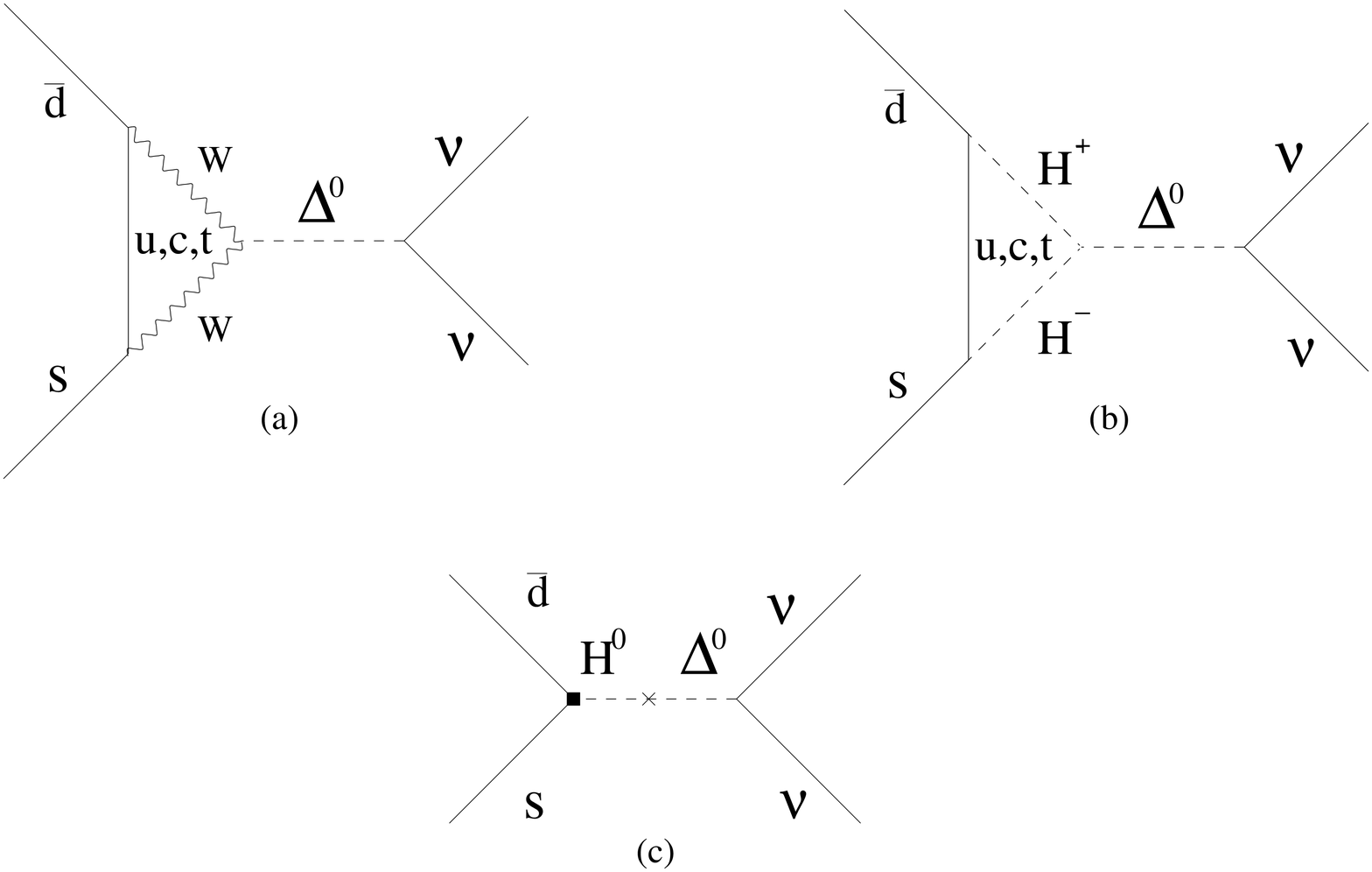,angle=0,width=13cm,height=12cm}}
\end{center}
\caption{CPC penguin diagrams mediated by a triplet Higgs.}
\label{figDel}
\end{figure}

The dominant new contributions to the $K_L\to\pi^0\nu\bar\nu$
decay come from the diagrams presented in
fig. \ref{figDel}, which generate the following effective
Hamiltonian:
\begin{equation}
{\mathcal{H}}^{\Delta}_{eff}=\sum_{\ell}\frac{G_{F}}{\sqrt{2}}
\frac{\alpha }{2\pi \sin ^{2}\Theta _{W}}
(\bar{d}s)({\nu }_{\ell } ^T i\tau ^2 \nu _{\ell })
(X_\ell^{(a)}+X_\ell^{(b)}
+X_\ell^{(c)})
\, ,
\label{Hpen}
\end{equation}
where the $X_\ell^{(i)}$ function corresponds to the
diagram in fig~\ref{figDel}(i).
In the $X_\ell^{(a)}$ and $X_\ell^{(b)}$
terms, the L breaking effects enter through, respectively,
the $\Delta WW$ and $\Delta HH$ couplings and
is related to the spontaneous breaking ($\vev{\Delta^0}\neq0$):
\begin{eqnarray}
X_\ell^{(a)}+{X_\ell}^{(b)}
&=&-m_{\nu }^\ell m_{s}\sum_{i}\lambda _{i}
\left(\frac{\sin^2 \gamma}{m_{\xi_1}^2}+
\frac{\cos^2 \gamma}{m_{\xi_2}^2}\right)
\left(\frac{1}{2}
+2x_i\frac{\lambda-\widetilde{\lambda}}{g^2}\right)
\nonumber \\
&\times& \left [1+ \frac{x_{i}}{(x_{i}-1)^3}
\left ( 2x_{i}\ln (x_{i})+1-x_{i}^{2} \right) \right]
\label{XPab} \, ,
\end{eqnarray}
where $x_{i}=(m_{i}/M_{W})^{2}$ and
$\lambda_{i}=V^{*}_{is}V_{id}$.
Since the top quark contribution is dominant,
(\ref{XPab}) can be simplified:
\begin{eqnarray}
X_\ell^{(a)}+X_\ell^{(b)}
&\approx&
-\lambda_t m_{\nu }^\ell m_{s}
\left(\frac{\sin^2 \gamma}{m_{\xi_1}^2}+
\frac{\cos^2 \gamma}{m_{\xi_2}^2}\right)
\nonumber \\
&\times&
x_t\left[\left(\frac{1}{2}
+2x_t\frac{\lambda-\widetilde{\lambda}}{g^2}\right)
\frac{\left( 2x_{t}\ln (x_{t})+1-x_{t}^{2}\right)}{(x_{t}-1)^3}
+2\frac{\lambda-\widetilde{\lambda}}{g^2}\right]
\, . \label{XPabGIM}
\end{eqnarray}
In the $X_\ell^{(c)}$ term,
the L breaking effect enters through the
$\Delta-H$ mixing and is related to both the soft breaking ($m\neq0$)
and the nonzero VEV of $\Delta^0$.
The calculation of $X_\ell^{(c)}$ is simplified by
the use of the $sdH$ effective coupling $\Gamma_{sdH}$,
[represented by a square in fig.~\ref{figDel}(c)]
which was calculated within the SM in ref.~\cite{WY}.
For our model $\Gamma_{sdH}$ should be expressed in terms of the appropriate
masses and mixing angles.
The mixing of the charged scalar can, however,
be neglected [see eq.~(\ref{tanet'})] and therefore we
use directly the SM calculation:
\begin{equation}
\Gamma_{sdH}=-\lambda _{t}\frac{g^{3}}{128\pi^{2}}
\frac{m_{t}^{2}m_{s}}{M_{W}^{3}} \left(\frac{3}{2}
+\frac{4\lambda_H}{g^2}f_2(x_t)
\right)(1+\gamma^5) \,,
\label{effSDH}
\end{equation}
where
\begin{equation}
f_2(x)= \frac{x}{2(1-x)^2}\left(-\frac{x}{1-x}\ln x
+\frac{2}{1-x}\ln x -\frac{1}{2}-\frac{3}{2x}\right)
\label{f_x} \, .
\end{equation}
Then $X_\ell^{(c)}$ is given by
\begin{eqnarray}
X_\ell^{(c)}=-\lambda _{t}m_\nu^\ell m_{s}
\frac{\sin2\gamma}{8}\frac{v_H}{v_\Delta}
\left(\frac{1}{m_{\xi_1}^2}-
\frac{1}{m_{\xi_2}^2}\right)
x_t \left(\frac{3}{2}
+\frac{4\lambda_H}{g^2}f_2(x_t)\right)
\label{XPc} \, .
\end{eqnarray}

Note that since the lepton and quark operators in the effective
Hamiltonian ${\mathcal{H}}^{\Delta}_{eff}$ in eq.~(\ref{Hpen}) are
scalar operators, the new contributions are CPC.

We are interested in finding the upper bound on the new contributions.
We therefore focus on the region in parameter space that maximizes them.
For all scalar masses, we will use a lower bound of 45 $GeV$, thus
avoiding any conflict with constraints from the invisible width of
the $Z$ boson. From eqs.~ (\ref{mim}) and (\ref{mcharge}) we then find:
\begin{equation}
{\rm sign}(m/v_\Delta)=-1
,\ \ \ |m/v_\Delta|\gtrsim 1
\label{lowerm} \, .
\end{equation}
Substituting the proper values for the masses and angles
(\ref{tang}), (\ref{mrHL}) into the functions
$X_\ell^{(a)},X_\ell^{(b)}$ (\ref{XPab})
and $X_\ell^{(c)}$ (\ref{XPc}) we find:
\begin{eqnarray}
X_\ell^{(a)}+X_\ell^{(b)} \sim
\lambda _{t}\frac{m_\nu^\ell m_{s}}{v_H^2}
\left(\frac{v_\Delta}{m}+O\left(v_\Delta^2/v_H^2\right)
\right)
\nonumber \\
X_\ell^{(c)}\sim
\lambda _{t}\frac{m_\nu^\ell m_{s}}{v_H^2}
\left(\frac{v_H^2}{\lambda_H v_H^2+mv_\Delta}
+O\left(v_\Delta^2/v_H^2\right)\right)
\label{estimXs}
\end{eqnarray}

For $m \gg v_H$
$X_\ell^{(a)}+X_\ell^{(b)}$ and
$X_\ell^{(c)}$
are all highly suppressed, together
with the total rate.
Larger contributions are found with
$m$ in the intermediate regime
$v_\Delta \lesssim m \lesssim v_H$, where
$X_\ell^{(a)}+X_\ell^{(b)}$ and
$X_\ell^{(c)}$ are all of the order of
$\frac{m_\nu^\ell m_s}{v_H^2} \sim
\frac{m_\nu^\ell m_{s}}{m^2_{\xi_{1,2}}}$.

The $\nu_m\nu_m\Delta$ coupling, that is $f_m$ of eq.
(\ref{tripmass}), is proportional to $m_{\nu_m}/v_\Delta$.
Naively, one may think that this mechanism of inducing
neutrino masses can induce a contribution to the
$K_L\to\pi\nu\nu$ rate that is enhanced by a factor of order
$v_H/v_\Delta$ compared to the mechanisms of \cite{GP}.
We find that
this is not the case. For the diagrams in fig. \ref{figDel}(a,b),
there is a $v_\Delta$-factor in the $WW\Delta$ and $HH\Delta$
couplings. For the diagram in fig. \ref{figDel}(c), there is a factor of
$\sin2\gamma v_H(1/m_{\xi_1}^2-1/m_{\xi_2}^2)\sim v_\Delta/v_H^2$.
In either case, the final result is proportional to $f_m v_\Delta
\sim m_{\nu_m}$ and there is no enhancement.

We are now in a position to compare the contribution of the
triplet scalar to the leading, CPV one \cite{IL}:
\begin{equation}
R_{CPV}^{\Delta}=\frac{\Gamma _{\Delta}
(K_{L} \rightarrow \pi ^{0} \nu \bar{\nu})}{\Gamma _{CPV}
(K_{L}\rightarrow \pi ^{0}\nu
\bar{\nu})}\lesssim \left(\frac{M_{K}}{m_{\xi_{light}}}
\frac{m_{\nu }}{m_{\xi_{light}}}\right)^2\approx 10^{-11}
\left[\frac{m_{\nu }}{10~  MeV}\right] ^{2}
\left[\frac{45~  GeV}{m_{\xi_{light}}}\right] ^{4}
\, ,  \label{RGtripCPCCPV}
\end{equation}
where $m_{\xi_{light}}$ stands for the lighter between $m_{\xi_1}$
and $m_{\xi_2}$.
While the direct bound is $m_{\nu_\tau}\leq 18.2\ MeV$\cite{EPJ},
there is a significantly stronger bound from cosmology,
$m_\nu\lesssim 10~eV$ \cite{HN}.
Therefore, very likely, $R^\Delta_{CPV}\lesssim 10^{-23}$.

%%%%%%%%%%%%%%%%%%%%%%%%%%%%%%%%%%%%%%%%%%%%%%%%%%%%%%%%%%%%
%%%%%%%Part Two%%%%%%%%%%%%%%%%%%%%%%%%%%%%%%%%%%%%%%%%%%%%%
%%%%%%%%%%%%%%%%%%%%%%%%%%%%%%%%%%%%%%%%%%%%%%%%%%%%%%%%%%%%

\section{The Zee Model}

\subsection{The Model}

The Zee model \cite{Zee} enlarges the SM scalar sector
by a charged $SU(2)$ singlet $\phi ^+$ and
an $SU(2)$ doublet $H_2$.
The SM doublet is denoted by $H_1$.
The Lagrangian of the model is \cite{Zee}:
\begin{eqnarray}
{\mathcal{L}}_{Zee} \equiv {\mathcal{L}}_{SM}+
f_{mn}\epsilon _{\alpha \beta}({L^{m\alpha}} ^T C{L^{n\beta}}_L)\phi^+
+\mu (H_1^{\dagger} H_2+H_2^{\dagger} H_1)
+V(H_1,H_2,\phi^+) \, ,
\label{LZ}
\end{eqnarray}
where $L$ is a lepton doublet, $\alpha,\beta$ are $SU(2)$ indices,
$m,n$ are flavor indices,
$C$ is the Dirac charge conjugation matrix and
$V(H_1,H_2,\phi^+)$ is the most general scalar potential which
respects the SM gauge symmetries and $L$~\cite{Santa}:
\begin{eqnarray}
V(H_1,H_2,\phi^+) &\equiv&
\lambda_1\left(\left|H_1\right|^2-\frac{a_1^2}{2}\right)^2
+\lambda_2\left(\left|H_2\right|^2 -\frac{a_2^2}{2}\right)^2
+\lambda_3\left(\left|H_1\right|^2 -\frac{a_1^2}{2}\right)
\left(\left|H_2\right|^2 -\frac{a_2^2}{2}\right)
\nonumber \\
&+&\lambda_4(\left|H_1\right|^2\left|H_2\right|^2
-{H_1}^\dagger H_2{H_2}^\dagger H_1 )
+\lambda_5 \left|{H_1}^T i\tau^2 H_2 \right|^2
+\lambda_6 m^2\left| \phi^+\right|^2
\nonumber \\
&+&\lambda_7 \left|\phi^+\right|^4
+\lambda_8 \left|\phi^+\right|^2\left|H_1\right|^2
+\lambda_9 \left|\phi^+\right|^2\left|H_2\right|^2
+M({H^1}^T i\tau^2 H^2\phi^- +h.c)
\, .
\label{Vscal}
\end{eqnarray}
Both Higgs doublets develop
nonzero VEVs:
\begin{equation}
\vev{{H^0}_{1,2}}\equiv \frac{v_{1,2}}{\sqrt{2}}
\, .
\label{VEVs}
\end{equation}
For $\mu=0$, the parameters $a_{1,2}$ would be equal to
$v_{1,2}$ respectively. With a nonzero $\mu$
term, $v_{1,2}$ are complicated functions of
$a_i$, $\lambda_i$ and $\mu$.
The $L$-charges of the scalars fields are:
\begin{equation}
L(H_1)=0,~
L(H_2)=2,~
L(\phi^+)=-2
\, .
\label{Lcharges}
\end{equation}
Then we see that $L$ is broken spontaneously via the
nonzero VEV of $H_2$.
The $\mu$-term in eq.~(\ref{LZ}) breaks $L$ explicitly
so that the phenomenologically
unacceptable Majoron~\cite{EPJ,Santa,LepV} is avoided.
For simplicity, we assume that CP is conserved in the lepton sector and
take all the dimensionless scalar couplings to be of order one.

In order to calculate the new contribution to the decay, the
scalar mass eigenstates and mixing angles should be
identified.
There are seven physical and three unphysical combinations
(eaten by the $Z$ and the $W^{\pm}$ bosons).
Out of the seven physical combinations, four are charged
and three are neutral.
The neutral ones are the would-be Majoron [which gains mass
due to the $\mu$ term in eq.~(\ref{LZ})] and two other real fields.
As long as $\mu$ is real, the real and
imaginary parts of $H_{1,2}^0$ remain unmixed
(as in the cases in which L is spontaneously broken
\cite{Santa,Pet}), thus the imaginary physical
combination corresponds to the would-be Majoron which is
irrelevant to our calculation.

The real neutral mass matrix is read from the
scalar potential of eq. (\ref{LZ}):
\begin{equation}
M_{\cal R} = \frac{1}{2}
\pmatrix{2\lambda_1(3{v_1}^2-{a_1}^2)+\lambda_3({v_2}^2-{a_2}^2) &
 2\lambda_{3} v_1 v_2 +\mu \cr
 2{\lambda_{3}} v_1 v_2 +\mu & 2\lambda_2(3{v_2}^2-{a_2}^2)
 +\lambda_3( {v_1}^2-{a_1}^2)}
 \, . \label{MH1H2}
\end{equation}
The eigenvectors of $M_{\cal R}$ are:
\begin{eqnarray}
\rho_1 &=& \Re (H_1^0)\cos\theta+\Re(H_2^0)\sin\theta
\nonumber \\
\rho_2 &=& -\Re(H_1^0)\sin\theta+\Re (H_2^0)\cos\theta
\, . \label{rhos}
\end{eqnarray}
Each eigenvector corresponds to an eigenvalue denoted by
$m_{\rho_{1,2}}$.
The masses $m_{\rho_{1,2}}$ and $\tan \theta$ are complicated
functions of $a_i$, $\lambda_i$ and $\mu$.

The charged mass matrix is:
\begin{equation}
M_{C} = \frac{1}{2}
\pmatrix{
\bar{\lambda}v_2^2-2\mu\frac{v_2}{v_1}&
-\bar{\lambda}v_1v_2 +2\mu& \sqrt{2}Mv_2 \cr
-\bar{\lambda}v_1v_2+2\mu &
\bar{\lambda}v_1^2-2\mu\frac{v_1}{v_2}& -\sqrt{2}Mv_1 \cr
\sqrt{2}Mv_2 &-\sqrt{2}Mv_1&
2\lambda_6 m^2+\lambda_8 v_1^2+\lambda_9 v_2^2}
 \, , \label{Mcharged}
\end{equation}
where $\bar{\lambda}\equiv \lambda_4+\lambda_5$.
The massless combination of the charged fields
(see e.g refs.~\cite{Haber,Hunter}) is:
\begin{equation}
G^{\pm}= H_1^{\pm}\cos\delta+H_2^{\pm}\sin\delta
 \, , \label{G+-}
\end{equation}
with
\begin{equation}
\tan\delta=\frac{v_2}{v_1}
 \, . \label{tandel}
\end{equation}
The massless combination is not affected by the addition of
the singlet field $\phi^+$, nor by the $\mu$ term,
since it is determined (according to the Goldstone theorem) only
by the  broken $SU(2)_L$ generators.
The physical charged fields are given by the following linear
combinations of $H_{1,2}^\pm$ and $\phi^\pm$
(which must be orthogonal to $G^{\pm}$):
\begin{eqnarray}
\chi_1^{\pm}&=& (-H_1^{\pm}\sin\delta
+H_2^{\pm}\cos\delta)\cos\beta+\phi^{\pm}\sin\beta
\nonumber \\
\chi_2^{\pm}&=& -(-H_1^{\pm}\sin\delta
+H_2^{\pm}\cos\delta)\sin\beta+\phi^{\pm}\cos\beta
\, , \label{chis}
\end{eqnarray}
with $\tan\beta$ being a complicated function of the scalar
potential parameters.
The new interactions in eq.~(\ref{LZ}) induced a neutrino
Majorana mass matrix $M^{\nu}_{m\ell}$.
For simplicity we assume the dominance of the
one loop induced mass \cite{Zee,Santa,Pet,Jar}:
\begin{equation}
M_{m \ell}^\nu = \frac{2\sqrt{2}}{(4\pi)^2}
f_{m \ell}\tan\delta \sin 2\beta
\frac{(m_{\ell}^2-m_m^2)}{M_W}
\ln\left(\frac{m_{\chi_2}}{m_{\chi_1}}\right)
\, ,  \label{mnuZ}
\end{equation}
with $m,\ell$ flavor indices.

\subsection{The $K_L\to\pi^0\nu\bar\nu$ Decay Rate}

\begin{figure}[hbct]
\begin{center}
\mbox{\epsfig{figure=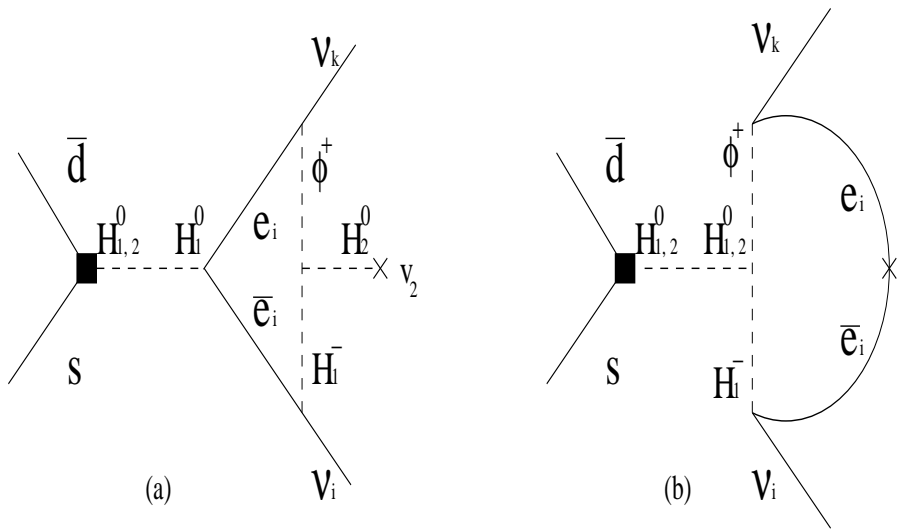,angle=0,width=12cm,height=7cm}}
\end{center}
\caption{CPC diagrams in the Zee model.}
\label{figZ}
\end{figure}

The new interactions also generate new contributions
to the $K_L \to \pi^0 \nu \bar{\nu}$ decay.
Note that the one loop induced mass and the new contributions
to the decay
(shown in fig.~\ref{figZ}) are related to mixing between the
charged scalars and
therefore vanish in the limit $\delta,\beta \to 0$.
Below we concentrate on the contributions which are dominant
when $\sin\delta$ is small. Contributions that are of higher order
do not modify significantly the results, even with $\sin\delta\sim1$,
and thus they are omitted. Note that, since $v_1$ induces the top mass,
we always have $\tan\delta\lesssim1$.

The scalar operator is induced by
the neutral Higgs mediated penguin diagram shown in
fig. \ref{figZ}~. The square in the figures
represents an effective $sdH_{1,2}$ vertex
denoted by $\Gamma_{sdH_{1,2}}$, similar to
the one we encountered in section II.B~.
The effective vertices  $\Gamma_{sdH_{1,2}}$
induced by the Zee model fields differ from the
one calculated within the SM~\cite{WY}. Since,
however, we are only interested in finding
an upper bound on the decay rate, we can simply
set the charged mixing angles factors to one,
and replace the boson masses in the propagator
(as explained in section II.B) with
$m_{\chi_{light}}\sim 45~GeV$. Then we get:
\begin{equation}
\Gamma_{sdH_{1,2}}\lesssim \frac{g^3 \lambda _{t}}{128\pi^{2}}
\frac{m_{t}^{2}m_{s}}{m_{\chi_{light}}^{3}}(1+\gamma^5)
\times[C_{sdH_{1}},C_{sdH_{2}}]
\label{effSDHZ} \, ,
\end{equation}
where $C_{sdH_{i}}$ is a constant of $O(10)$.~Neglecting 
subdominant contributions of~$O[(m_\ell/m_{\rho_i,\chi_i})^4]$,
the diagrams in fig.~\ref{figZ} generate the
following CPC effective Hamiltonian:
\begin{equation}
{\mathcal{H}}^{\phi}_{eff}=\frac{G_{F}}{\sqrt{2}}
\frac{\alpha}{2\pi \sin ^{2}\Theta _{W}}\lambda _{t}
(\bar{d}s)({\nu }_{m} ^T i\sigma ^2  \nu _{L \ell})
(Y_{\ell m}^{(a)}+Y_{\ell m}^{(b)})
\, ,  \label{HZ}
\end{equation}
where the $Y_{\ell m}^{(i)}$ function corresponds to the
diagram in fig.~\ref{figZ}(i). $Y_{\ell m}^{(a)}$
is given by:
\begin{eqnarray}
Y_{\ell m}^{(a)}&=&2\frac{M^{\nu}_{\ell m}M_W}
{\cos\delta}
\frac{m_{t}^{2}m_{s}}{m_{\chi_{light}}^{3}}
\left[C_{sdH_{1}}\left(\frac{\cos^2\theta}{m_{\rho_1}^2}+
\frac{\sin^2\theta}{m_{\rho_2}^2}\right)
+C_{sdH_{2}}\frac{\sin 2 \theta}{2} \left(\frac{1}{m_{\rho_1}^2}-
\frac{1}{m_{\rho_2}^2}\right)\right]
\label{XZa} \, .
\end{eqnarray}
The calculation of the diagram in fig.~\ref{figZ}(b)
is much more involved.
This is due to the fact that each of the trilinear
couplings $H^0_i H^\pm_j H^\mp_k$ and $H^0_i H^\pm_j \phi^\mp$
($i,j,k=1,2$) translates into a set of eight couplings in the
mass basis. In order to estimate the upper bound on the rate we set
(again) all mixing factors to unity and replace the charged
scalar propagators by their maximal values,
i.e. $\frac{1}{k^2-m_{\chi_1}^2}
-\frac{1}{k^2-m_{\chi_2}^2} \to \frac{1}{k^2-m_{\chi_{light}}^2}$
and $\frac{1}{k^2-m_{\chi_1}^2}+\frac{1}{k^2-m_{\chi_2}^2}
\to \frac{2}{k^2-m_{\chi_{light}}^2}$~.
We then find:
\begin{eqnarray}
Y_{\ell m}^{(b)}& \lesssim&
\frac{\sqrt{2}}{g64\pi^2}f_{\ell m}(m_{\ell}^2-m_m^2)
\frac{M_W^2}{m_{\rho_1}^2}
\frac{ m_{t}^{2}m_{s}}{m_{\chi_{light}}^5}
\left[F_1(\theta,m_{\rho_i},m_{\chi_i},M)+
F_2(\theta,m_{\rho_i},m_{\chi_i},M) \right]
\nonumber \\
&=&
\frac{1}{8g}\frac{\cot\delta}{\sin 2\beta} 
\left(\ln\frac{m_{\chi_2}}{m_{\chi_1}}\right)^{-1}
\frac{M_{m \ell}^\nu m_{s}}{m_{\rho_1}^2}
\frac{ m_{t}^{2}M_W^3}{m_{\chi_{light}}^5}
\nonumber \\
&\times&
\left[F_1(\theta,m_{\rho_i},m_{\chi_i},M)+
F_2(\theta,m_{\rho_i},m_{\chi_i},M) \right]
\label{XZb} \, ,
\end{eqnarray}
where
\begin{eqnarray}
F_1(\theta,m_{\rho_i},m_{\chi_i},M)&=&
\left[C_{sdH_{1}}\left(\cos^2\theta+
\sin^2 \theta\frac{ m_{\rho_1}^2}{m_{\rho_2}^2}\right)
+C_{sdH_2}\frac{\sin 2 \theta}{2}
\left(-1+\frac{m_{\rho_1}^2}{m_{\rho_2}^2}\right)\right]
\nonumber \\
&\times&
\left(2\lambda_1+\frac{1}{2}\bar{\lambda}+2\lambda_3+2\lambda_8
+3\frac{M}{v_1} \right)
\label{F1} \, ,
\end{eqnarray}
and
\begin{eqnarray}
F_2(\theta,m_{\rho_i},m_{\chi_i},M)&=&
\left[C_{sdH_{2}}\left(\sin^2\theta+
\cos^2 \theta\frac{ m_{\rho_1}^2}{m_{\rho_2}^2}\right)
-C_{sdH_1}\frac{\sin 2 \theta}{2}
\left(-1+\frac{m_{\rho_1}^2}{m_{\rho_2}^2}\right)\right]
\nonumber \\
&\times&
\left(2\lambda_2+2\lambda_3+\frac{1}{2}\bar{\lambda}
+2\lambda_9
-5\frac{M}{v_1} \right)
\label{F2} \, .
\end{eqnarray}
The $M$-term in the scalar potential (\ref{Vscal})
does not break any symmetry. One might think then that the ratio
$\frac{M}{v_1}$ that appears in eqs.~(\ref{F1}) can be arbitrarily
large and enhance the decay rate. This is however not the case.
For very large $M$, the scalar potential would spontaneously break
$U(1)_{EM}$. An even stronger upper bound on $M$ comes from two loop
contributions to the neutrino masses. We assumed, for simplicity, that
the neutrino masses are dominated by the one loop contributions.
This assumption requires that $\frac{M}{v_1}$ is not much larger
than the various $\lambda$ couplings.
Adding (\ref{XZa}) to (\ref{XZb}), and
applying the approximation described above eq.~(\ref{XZb}) we find:
\begin{eqnarray}
Y_{\ell m}^{(a)}+Y_{\ell m}^{(b)}
&\lesssim&
\frac{m_{\nu_3} M_W}{m^2_{\rho_{light}}}
\frac{m_{t}^{2}m_{s}}{m_{\chi_{light}}^{3}}
\left[C_{sdH_{1}}\left(4+3\frac{M_W^2}{m_{\chi_{light}}^2}\right)
+C_{sdH_{2}}\left(1+3\frac{M_W^2}{m_{\chi_{light}}^2}\right)\right]
\nonumber \\
&\lesssim&
10^2\frac{m_{\nu_3} M_W}{m^2_{\rho_{light}}}
\frac{m_t^2 m_{s}}{m_{\chi_{light}}^{3}}
\label{XZab} \, ,
\end{eqnarray}
where $m_{\nu_3}$ is the largest eigenvalue of $M^{\nu}_{\ell m}$.

We can now compare the CPC rate of the Zee model with the
leading CPV one \cite{IL}:
\begin{eqnarray}
R_{CPV}^{Zee}=\frac{\Gamma _{Zee}
(K_{L} \rightarrow \pi ^{0} \nu \bar{\nu})}{\Gamma _{CPV}
(K_{L} \rightarrow \pi ^{0} \nu
\bar{\nu})}
&\lesssim&
\left(10^2\frac{m_\nu M_W}{m^2_{\rho_{light}}}
\frac{m_{t}^{2}m_K}{m_{\chi_{light}}^{3}}\right)^2
\nonumber \\
&\approx&
10^{-5}\left[ \frac{m_{\nu }}{10~  MeV}
\left(\frac{45~  GeV} {m_{\rho_{light}}}\right) ^2
\left(\frac{45~  GeV} {m_{\chi_{light}}}\right) ^3
\right] ^{2}
\,.
\label{RGZCPCCPV}
\end{eqnarray}
Since, very likely, $m_{\nu _{\tau }}\leq 10~eV$ \cite{Neu} ,
we expect $R _{CPV}^{\Delta}\lesssim 10^{-17}$.
Recall that this upper bound was obtained using
some crude approximations and is expected to be even smaller
for an exact calculation.

\section{Final Conclusions}

In this work we examined the question of whether the SM CP violating
contribution to the $K_{L}\rightarrow \pi ^{0}\nu \bar{\nu}$ decay
is still dominant in the presence of new scalars that induce Majorana
masses for neutrinos. We found the following unambiguous answers:
\begin{itemize}
\item[(i)] For CPC contributions induced by an $SU(2)$-triplet
scalar, we get:
\begin{equation}
\frac{\Gamma _{\Delta} (K_{L} \rightarrow
\pi ^{0} \nu \bar{\nu})}{\Gamma _{SM}(K_{L}
\rightarrow \pi ^{0} \nu \bar{\nu})}\lesssim
\left(\frac{M_{K} m_{\nu }}
{m_{\xi_{light}}^2}\right)^{2}
\lesssim 10^{-11}
~ .  \label{Rtrip}
\end{equation}

\item[(ii)] For CPC contributions that are generated in the
Zee model, we get:
\begin{equation}
\frac{\Gamma _{Zee}(K_{L}\rightarrow
\pi ^{0} \nu \bar{\nu})}{\Gamma _{SM}(K_{L}
\rightarrow \pi ^{0}\nu \bar{\nu})}
\lesssim
\left(10^2\frac{m_\nu M_W}{m^2_{\rho_{light}}}
\frac{m_{t}^{2}m_K}{m_{\chi_{light}}^{3}}\right)^2
\lesssim
10^{-5}
\, .
\label{RZ}
\end{equation}
\end{itemize}
In obtaining the final bounds in eqs. (\ref{Rtrip}) and (\ref{RZ})
we used the direct upper bound on $m_{\nu_\tau}$ of order 10 MeV.
If we use the cosmological bound, $m_\nu\lesssim10\;eV,$ then the
bounds become stronger by twelve orders of magnitude.
It is clear then that the $K_{L}\rightarrow \pi^0\nu \bar{\nu}$ decay
provides a very clean measurement of fundamental, CP violating
properties and that it does not probe neutrino masses.

\vspace{3mm}
{\large Acknowledgments} \vspace{3mm}

I thank Yossi Nir for his guidance in this work.
I also thank Sven Bergman and Galit Eyal for fruitful discussions.

\end{document}